\documentclass[aps,twocolumn,showpacs,amsmath,amssymb,amsfonts]{revtex4}
\usepackage{graphicx}
\usepackage{bm}
\usepackage{dcolumn}
\newcommand{\tst}{\textstyle}

\newcommand{\mbf}{\mathbf}
\newcommand{\mrm}{\mathrm}
\newcommand{\ud}{\mathrm{d}}
\begin{document}

\author{Zbigniew Idziaszek}
\affiliation{Institute of Theoretical Physics, University of Warsaw, Ho{\.z}a 69,
00-681 Warsaw, Poland}
%\affiliation{Center for Theoretical Physics, Polish Academy of Sciences, Aleja Lotnik{\'o}w 32/46, 02-668 Warsaw, Poland}
\title{Analytical solutions for two atoms in a harmonic trap: $p$-wave interactions}

\begin{abstract}
We derive analytical solutions for the system of two ultracold spin-polarized fermions interacting in $p$ wave and confined in an axially symmetric harmonic trap. To this end we utilize $p$-wave pseudopotential with an energy-dependent scattering volume. This allows to describe the scattering in tight trapping potentials in the presence of scattering resonances. We verify predictions of the pseudopotential treatment for some model interaction potential, obtaining an excellent agreement with exact energy levels. Then we turn to the experimentally relevant case of neutral atom interactions in the vicinity of a $p$-wave Feshbach resonance. In the framework of the multichannel quantum-defect theory we derive relatively simple formula for an energy-dependent scattering volume, and later we apply it to investigate the energy spectrum of trapped atoms close to the $p$-wave Feshbach resonance.
\end{abstract}

\pacs{34.50.Cx, 37.10.Gh, 03.65.Ge}

\maketitle

\section{Introduction}

In the regime of ultracold temperatures atomic collisions are typically dominated by $s$-wave scattering. The situation changes in the presence of scattering resonances, which can strongly enhance the contribution from higher partial waves. This has been demonstrated in experiments by employing Feshbach resonances to enhance $p$-wave interactions in Fermi gases \cite{Regal,Zhang,Schunck,Gunter} and Fermi-Bose mixtures \cite{Ospelkaus}. So far realization of $p$-wave superfluids is hindered by the presence of strong collisional losses in the vicinity of resonances, and, as a consequence, short lifetime of the ultracold gas. Similar losses affect the lifetime of $p$-wave Feshbach molecules, that has been recently created \cite{Gaebler,Fuchs,Inada} and theoretically investigated \cite{Gubbels}. Realization of $p$-wave superfluids is of particular concern because of their very interesting properties, e.g. rich phase diagram with a variety of classical and quantum phase transitions \cite{Ho,Ohashi,Gurarie,Cheng,Iskin}.

Theoretical description of ultracold gases typically assumes modeling of an atom-atom interaction in terms of a contact potential. In the case of $s$-wave interactions this is typically done by means of Fermi pseudopotential \cite{Fermi,Huang}. Analogous pseudopotential for higher partial waves has been developed several years ago by Huang and Yang \cite{Huang,HuangBook}, however, their original derivation turned out to be incorrect \cite{Roth,Derevianko} and resulting pseudopotential valid only for $s$-wave scattering. Discovery of an error in the derivation of Huang and Yang has been a motivation for a series of papers proposing several forms of pseudopotentials for higher partial waves \cite{Roth,Stock,Kanjilal,Pricoupenko,Derevianko,IdziaszekPRL,Reichenbach,Stampfer}. Such pseudopotentials have been applied, for instance, to confirm the existence of geometric resonances \cite{Olshanii,Granger} in the scattering of fermions in quasi-two-dimensional systems \cite{IdziaszekPRL} or to investigate trap-induced shape resonances of $p$-wave interacting atoms \cite{Reichenbach}.

In this paper we apply the pseudopotential method to develop analytical solutions for two spin-polarized fermions interacting in $p$-wave and confined in the axially symmetric harmonic trap. Some of the results for pancake-shape traps have been already presented in our earlier work \cite{IdziaszekPRL}. Here we present much more detailed analysis, including the analytical formulas for traps of arbitrary aspect ratio. We verify the accuracy of the pseudopotential treatment by comparison with exact energy levels for a model square-well potential. Finally we consider a relevant case of atom-atom interactions tuned by the $p$-wave Feshbach resonances. Our results are of direct interest for studies of spin-polarized Fermi gases in optical lattices \cite{Gunter}, where tightly confined pairs of atoms can be obtained and $p$-wave Feshbach molecules can be created by applying time-dependent magnetic fields, similarly to $s$-wave Feshbach resonances \cite{Stoferle,Ospelkaus:2006,Diener}.

An important part of our analysis constitute rigorous derivation of pseudopotentials for higher partial waves. Here, for $p$-wave interactions we utilize slightly modified version of zero-range potential than introduced in \cite{IdziaszekPRL}, which is more convenient for our analytical calculations in the trap, and allows for simple generalization for partial waves $\ell >1$. Because of the variety of pseudopotentials in the literature, we devote a small section on comparison of our pseudopotential to other formulations, discussing the conditions when they become equivalent.

In contrast to $s$-wave interaction where the scattering amplitude acquires a constant value at low energy, and can be characterized with a single parameter (e.g. $s$-wave scattering length), pseudopotentials  in higher partial requires inclusion of an energy-dependent amplitude for modeling of resonant interactions. Existence of the centrifugal barrier makes the higher partial wave resonances intrinsically narrow and sensitive to the energy of the colliding particles. In our approach we employ multichannel-quantum defect theory (MQDT) for ultracold collisions \cite{Seaton,Greene,Mies}, that allows for an exact treatment of two-body scattering in the presence resonances and yields relatively simple expression for the energy-dependent amplitude of the pseudopotential due to the $p$-wave Feshbach resonance.

The structure of the paper is as follows. In section \ref{Sec:CorrPseud} we present detailed derivation of the pseudopotential for all partial wave interactions. Section \ref{Sec:OtherRepr} contains a comparison of different pseudopotentials proposed in the literature. Section \ref{Sec:PseudoSingle} presents simpler form of pseudopotential, that can be used when interaction takes place only in a single partial wave. Reader not interested in details of derivation, may skip sections \ref{Sec:CorrPseud}-\ref{Sec:PseudoSingle} and jump directly to section \ref{Sec:PseudoPD} where we introduce $p$-wave pseudopotential used throughout this paper, and also, for the sake of illustration, a pseudopotential for $d$-wave interactions. Section \ref{Sec:TwoAtoms} is devoted to analysis of two interacting spin-polarized fermions in a harmonic trap. In Section \ref{Sec:AnalRes} we derive some general analytical result for the energy spectrum. Specifying the form of the energy-dependent scattering volume, we apply these results to investigate the energy spectrum for a model square-well interaction (Section \ref{Sec:SquareWell}) and for an atom-atom interaction tuned with a magnetic Feshbach resonance (Section \ref{Sec:Feshbach}). We summarize in Section \ref{Sec:Summary} presenting some conclusions. Finally Appendices \ref{Sec:App1}-\ref{Sec:App3} present some technical details related to derivation of pseudopotentials.

\section{Pseudopotentials for higher partial wave interactions}

\subsection{Corrected pseudopotential of Huang and Yang}
\label{Sec:CorrPseud}

In this section we present detailed derivation of the generalized pseudopotential for all partial waves, introduced in Refs.~\cite{Derevianko,IdziaszekPRL}. It can be obtained in a manner similar to the original derivation of Huang and Yang \cite{Huang}, where the only difference is the rigorous treatment of the delta-function derivatives within the distribution theory. We start from the Schr\"odinger equation in the relative coordinate
\begin{equation}
\label{SchrEq}
\frac{\hbar^2}{2 \mu} (\Delta + k^2) \Psi( \mbf{r}) = V(r) \Psi( \mbf{r}),
\end{equation}
where $k^2 = 2 \mu E/ \hbar^2$, and $\mu$ denotes the reduced mass. We assume that potential $V(r)$ is central and has a finite range \footnote{More precisely, it is sufficient that $V(r)$ decays faster at infinity than $r^{-2}$, i.e. $\lim_{r \rightarrow \infty} V(r) r^2  = 0$}. Outside the range of potential, the wave function exhibits the following asymptotic behavior
\begin{equation}
\Psi_{a}(\mbf{r}) = \sum_{l=0}^{\infty} \sum_{m=-l}^{l} R_{l}(r) Y_{lm}(\mbf{\hat{r}}),
\end{equation}
Here, $Y_{lm}(\mbf{\hat{r}})$ are spherical harmonics, the radial wave functions $R_{l}(r)= C_{lm} (j_l(kr) - \tan \delta_l n_l(k r))$ are linear combinations of spherical Bessel and von Neumann functions $j_l(r)$ and $n_l(r)$, respectively, $C_{lm}$ are coefficients that depend on the boundary conditions at $r \rightarrow \infty$, and the phase shifts
$\delta_l$ are determined by the potential $V(r)$.
Our goal is to replace $V(r)$ by zero-range potential $V_\mrm{ps}(r)$, which
acts only at $\mbf{r}=0$ and gives the same asymptotic function $\Psi_{a}(\mbf{r})$, as the real potential $V(r)$. Following Huang and Yang \cite{Huang}, we investigate the behavior of the l.h.s. of Eq.~(\ref{SchrEq}) at $r \rightarrow 0$, substituting $\Psi = \Psi_a$.
Since $j_{l}(k r)$ is regular at $\mbf{r}=0$, the only
contribution to the pseudopotential comes from $n_l(k r)$ that is singular at small $r$: $n_l(x) = - (2l-1)!!/x^{l+1} + {\cal O}(1/x^{l-1})$. We have
\begin{align}
\label{SchrEq2}
(\Delta + k^2) \Psi_{a}(\mbf{r}) = & - \sum_{l=0}^{\infty} \sum_{m=-l}^{l} C_{lm} \tan \delta_l
Y_{lm}(\mbf{\hat{r}}) \nonumber \\
& \times \! \left( \frac{d^2}{d r^2} + \frac{2}{r} \frac{d}{d r} - \frac{l(l+1)}{r^2} + k^2 \right)\!n_l(kr)
\end{align}
For terms of the order of $1/r^{l-1}$ and higher in the expansion of $n_l(kr)$, we observe that the radial part of the Laplace operator produces terms that are exactly canceled by terms obtained from the multiplication by $k^2$. On the other hand, the most singular term $1/r^{l+1}$ gives no
contribution at $r \neq 0$, and the only nontrivial part comes from the radial part of the Laplace operator applied to $1/r^{l+1}$ at $r=0$. The latter can be calculated on the ground of the theory of generalized functions.

In our derivation we adopt the Hadamard finite part regularization of the singular function $1/r^{k}$, which can be defined as \cite{Estrada,Kanwal}
\begin{equation}
\label{Pf}
\begin{split}
\textrm{Pf} \left(\frac{1}{r^k}\right) = \lim_{\varepsilon \rightarrow 0} \Bigg[ & \frac{H(r-\varepsilon)}{r^k}
- \sum_{j=0}^{k-2} \frac{(-1)^j \delta^{(j)}(r)}{j! (k-j-1) \varepsilon^{k-j-1}}\\
& +  \frac{(-1)^{k-1} \ln \varepsilon \delta^{(k-1)}(r)}{(k-1)!} \Bigg]
\end{split}
\end{equation}
for $k$ integer and positive. This definition follows simply from substraction of diverging terms in the integral of the type $\int_\varepsilon^{\infty} \ud x \, \phi(x)/x^k$, $\varepsilon \rightarrow 0$ (see Appendix A for details).
The presence of the Heaviside function $H(r)$ is related to the fact that the radial wave
function is defined for $r \geq 0$. We stress that regularization procedure is not unique (see for instance \cite{Estrada}), however all regularizations must give the same results when integrated with a test function behaving like $r^k$ for small $r$. Regularization (\ref{Pf}) has the following properties \cite{Estrada}
\begin{align}
\label{Pf1}
r^q \textrm{Pf} \left(\frac{1}{r^k}\right) & = \textrm{Pf} \left(\frac{1}{r^{k-q}}\right),
\quad q=0,1,\ldots,k-1 \\
\label{Pf2}\
\frac{\bar{d}}{dr}
\textrm{Pf} \left(\frac{1}{r^k}\right) & = - k\textrm{Pf} \left(\frac{1}{r^{k+1}}\right) +
\frac{(-1)^k \delta^{(k)}(r)}{k!}
\end{align}
where $\bar{d}/(dr)$ represents generalized, distributio\-nal de\-ri\-vative, and $\delta^{(n)}(r)$ denotes the $n$-th derivative od the delta function. When the distribution $\delta^{(n)}$ is applied to a test function $\phi(r)$ that vanishes at $r \rightarrow 0$ as $r^{\alpha}$ with $\alpha \geq 1$, it yields
\begin{equation}
\label{deltan}
\langle \phi, \delta^{(k)}(r) \rangle = - k \langle \phi, \delta^{(k-1)}(r)/r \rangle, \quad
\lim_{r \rightarrow 0} \frac{\phi(r)}{r} < \infty,
\end{equation}
which follows easily from integration by parts. Combining \eqref{Pf1}, \eqref{Pf2} and \eqref{deltan} we obtain
\begin{equation}
\label{RadialPf}
\begin{split}
\left(\frac{\bar{d}^2}{dr^2} + \frac{2}{r} \frac{\bar{d}}{dr} - \frac{l(l+1)}{r^2} \right) &
\textrm{Pf} \left(\frac{1}{r^{l+1}}\right) \\
& = \frac{ (-1)^{l+1} (2l+1)}{l!} \frac{\delta^{(l)}(r)}{r^2}.
\end{split}
\end{equation}
By applying \eqref{deltan} twice we have extracted $1/r^2$ term, which is convenient when calculating matrix elements in spherical coordinates, because of the additional $r^2$ factor arising from the Jacobian. Finally, we express the coefficients $C_{lm}$ in terms of the radial wave function: $C_{lm} = ((2l+1)!!/[k^l (2l+1)!] (\partial_r^{2l+1} r^{l+1} R_l(r))_{\mbf{r}=0}$. The regularization operator $\partial_r^{2l+1} r^{l+1}$ removes the singular contribution from $n_l(kr)$, whereas the prefactor is determined by the small argument behavior of the $j_l(kr)$: $j_l(x) = x^l/(2l+1)!! + {\cal O}(x^{l+2})$. This leads to the pseudopotential \cite{IdziaszekPRL}
\begin{align}
\label{Vps}
V_\mrm{ps}(\mbf{r}) \Psi(\mbf{r}) = & \sum_{l=0}^{\infty}
\frac{(-1)^{l}\,[(2l+1)!!]^2}{4 \pi (2l)!\, l!} g_l
\frac{ \delta^{(l)}(r)}{r^2} \\
& \times  \left[
\frac{\partial^{2l+1}}{\partial {r^\prime}^{2l+1}} {r^\prime}^{l+1}
\!\int \! \! \ud\Omega^{\prime} P_l(\mbf{\hat{r}} \! \cdot \!\mbf{\hat{r}}^\prime)  \Psi(\mbf{r}^\prime)
\right]_{\mbf{r}^\prime=0} \nonumber,
\end{align}
where the angular integral over $\Omega^{\prime}$ acts a projection operator
on a state with a given quantum number $l$, and we have performed the summation over $m$ resulting  in the Legendre polynomial $P_l$. Here, $g_l = \hbar^2 a_l(k)^{2l+1}/(2 \mu) $, and $a_l(k)$ denotes the energy-dependent scattering length for partial wave $l$: $a_l(k)^{2l+1} = - \tan \delta_l(k)/k^{2l+1}$ \cite{Blume,Bolda}.
We observe that $l=0$ component yields the familiar $s$-wave contact potential
$2 \pi \hbar^2 a_0(k)/\mu \delta(\mbf{r})\partial_r r$, whereas
components with $l>0$ differ from the result of Huang and Yang
by a prefactor $(2l+1)/(l+1)$ \cite{HuangBook}.
%We note that in calculation of the matrix elements of $V_{ps}(\mbf{r})$ the differentiation of delta function can be replaced by differentiation of the function that acts on the l.h.s. of the pseudopotential, with a proper change of sign. Function acting on thr r.h.s. of the pseudopotential contributes only to the angular part of the integration.

Finally we note that for $l>1$ the Bessel function $n_l(x)$, apart from
the leading order singularity $1/x^{l+1}$, contains also lower order singularities, which assuming our regularization, in principle give rise to the lower-order derivatives of the delta distribution for $l$-th component of the pseudopotential. Such terms, however, do not give any contribution for functions behaving like $r^{l}$ for small $r$, and they do not appear when the other regularizations \cite{Stampfer} are applied.

\subsection{Other representations of pseudopotential}
\label{Sec:OtherRepr}

Now we show, the equivalence of \eqref{Vps} to other zero-range potentials derived recently in the literature \cite{Stock,Derevianko,Stampfer}. To this end we evaluate matrix elements of $V_\mrm{ps}$, making some assumption about the form of the wave function $\Phi(\mbf{r})$ acting to the l.h.s. of \eqref{Vps}. First, assuming that the $l$-wave components $\phi_l(r)$ of $\Phi(\mbf{r})$ behave as $\phi_l(r) \sim r^{\alpha}$ with $\alpha \geq l$ at small $r$, we can apply \eqref{deltan} recursively, to obtain
\begin{equation}
\label{deltal}
\langle \phi_l, \delta^{(l)}(r) \rangle = (-1)^l l! \langle \phi_l, \delta(r)/r^l \rangle, \quad
\lim_{r \rightarrow 0} \phi_l(r) \sim r^l.
\end{equation}
This combined with the identity $\delta (\mbf{r}) = \delta(r)/(4 \pi r^2)$ results in the pseudopotential \cite{Derevianko}
\begin{align}
\label{VpsSt}
\widetilde{V}_\mrm{ps}(\mbf{r}) \Psi(\mbf{r}) = & \sum_{l=0}^{\infty}
\frac{[(2l+1)!!]^2}{4 \pi(2l)!} g_l
\frac{\delta(r)}{r^{l+2}} \\
& \times  \left[
\frac{\partial^{2l+1}}{\partial {r^\prime}^{2l+1}} {r^\prime}^{l+1}
\!\int \! \! \ud\Omega^{\prime} P_l(\mbf{\hat{r}} \! \cdot \!\mbf{\hat{r}}^\prime)  \Psi(\mbf{r}^\prime)
\right]_{\mbf{r}^\prime=0} \nonumber,
\end{align}
The delta-shell pseudopotential of Stock {\it et al.} \cite{Stock}, differs from the latter representation by use of the delta-shell distribution $\lim_{s \rightarrow 0} \delta(r-s)/s^2$, instead of $\delta(r)/r^2$.

Another useful representation of the pseudopotential was introduced by Stampfer and Wagner \cite{Stampfer}
\begin{multline}
\label{VpsStamp}
\overline{V}_\mrm{ps}(\mbf{r}) \Psi(\mbf{r}) = \sum_{l=0}^{\infty} \sum_{m=-l}^{l}
\frac{4 \pi (-1)^{l}}{(2l)!!} g_l Y_{lm}(\partial)\delta(\mbf{r}) \\
\times  \left[
\frac{\partial^{2l+1}}{\partial {r^\prime}^{2l+1}} {r^\prime}^{l+1}
\!\int \! \! \ud\Omega^{\prime} Y_{lm}^{\ast} (\mbf{\hat{r}}^\prime) \Psi(\mbf{r}^\prime)
\right]_{\mbf{r}^\prime=0},
\end{multline}
where the partial differential operator $Y_{lm}(\partial)$ is obtained from the harmonic polynomial $r^l Y_{lm}(\mbf{\hat{r}})$, by replacing the Cartesian coordinates $x_k$ with the partial derivatives $\partial_{x_k}$. The operator $Y_{lm}(\partial)$, introduced by Maxwell \cite{SphericalHarmonics}, has the following properties
\begin{align}
Y_{lm}(\partial)r^{\lambda} = \lambda (\lambda-2) \ldots (\lambda-2l+2) r^{\lambda-l} Y_{lm}(\mbf{\hat{r}}), \\
\left.Y_{lm}^\ast(\partial) r^{l^\prime} Y_{l^\prime m^\prime}(\mbf{\hat{r}}) \right|_{\mbf{r}=0}  = \frac{(2l+1)!!}{4 \pi} \delta_{ll^\prime}\delta_{mm^\prime}
\label{OrthoProperty}
\end{align}
The equivalence of pseudopotentials \eqref{Vps} and \eqref{VpsStamp} follows from the following identity
\begin{equation}
\label{Ident1}
\frac{ \delta^{(l)}(r)}{r^2} Y_{lm}(\mbf{\hat{r}}) = \frac{4 \pi l!}{(2l+1)!!} Y_{lm}(\partial) \delta(\mbf{r}),
\end{equation}
which we prove in the Appendix B.

\subsection{Pseudopotentials for interactions in a single partial wave}
\label{Sec:PseudoSingle}

Quite often the atom-atom interaction occurs dominantly in a single partial wave, e.g. in the vicinity of resonances. In such cases we can consider only a single partial wave, and derive simpler form of the pseudopotential. To this end, we express the projection operator on a partial wave $l$, that appears int the square brackets in Eq.~\eqref{VpsStamp}, in terms of the differential operator $Y_{lm}(\partial)$. We observe that the only non vanishing contribution to
$\left[\partial_r^{2l+1} {r^\prime}^{l+1} \!\int \! \! \ud\Omega^{\prime} Y_{lm}^{\ast} (\mbf{\hat{r}}^\prime) \Psi(\mbf{r}^\prime)\right]_{\mbf{r}^\prime=0}$, comes from the terms of the symmetry $r^l Y_{lm}(\mbf{\hat{r}}^\prime)$. Therefore we can rewrite the regularization part of the pseudopotential, introducing an additional differentiation and multiplication by $r^\prime$,
\begin{multline}
\left[
\frac{\partial^{2l+1}}{\partial {r^\prime}^{2l+1}} {r^\prime}^{l+1}
\!\int \! \! \ud\Omega^{\prime} Y_{lm}^{\ast} (\mbf{\hat{r}}^\prime) \Psi(\mbf{r}^\prime)
\right]_{\mbf{r}^\prime=0} \\
 = \left[
\frac{\partial^{2l+1}}{\partial {r^\prime}^{2l+1}} {r^\prime}^{2l+1} \frac{1}{l!}
\frac{\partial^{l}}{\partial {r^\prime}^l}
\!\int \! \! \ud\Omega^{\prime} Y_{lm}^{\ast} (\mbf{\hat{r}}^\prime) \Psi(\mbf{r}^\prime)
\right]_{\mbf{r}^\prime=0}
\end{multline}
In a similar fashion we can transform the projection operator
\begin{equation}
\frac{1}{l!}
\frac{\partial^{l}}{\partial {r^\prime}^l}
\!\int \! \! \ud\Omega^{\prime} Y_{lm}^{\ast} (\mbf{\hat{r}}^\prime) \Psi(\mbf{r}^\prime) \rightarrow \frac{4 \pi}{(2l+1)!!} Y_{lm}(\partial^\prime)  \Psi(\mbf{r}^\prime),
\end{equation}
which does not affect the part of the wave function proportional to $(r^\prime)^l Y_{lm}(\mbf{\hat{r}}^\prime)$, as can be easily verified applying the property \eqref{Ident1}. This yields the following representation of the zero-range potential for partial wave $l$
\begin{multline}
\label{VpsStamp1}
V_l(\mbf{r}) \Psi(\mbf{r}) = \sum_{m=-l}^{l}
\frac{16 \pi^2 (-1)^{l}}{(2l+1)!} g_l Y_{lm}(\partial)\delta(\mbf{r}) \\
\times \left[
\frac{\partial^{2l+1}}{\partial {r^\prime}^{2l+1}} {r^\prime}^{2l+1} Y_{lm}^\ast(\partial^\prime)
\Psi(\mbf{r}^\prime) \right]_{\mbf{r}^\prime=0}
\end{multline}
Pseudopotential \eqref{VpsStamp1} involves only differential operators, that are typically much easier to handle than the angular integration projecting on a given partial wave $l$ in \eqref{VpsStamp}. We note that representation \eqref{VpsStamp1} is equivalent to \eqref{VpsStamp} only if the interactions occur in a single partial wave $l$, and the resulting wave function exhibits $1/r^l$ or $r^l$ behavior at small distances.

We observe that the summation over quantum number $m$ involves only the  differential operators $Y_{lm}(\partial)$. The sum over $m$ can be performed using an identity that directly follows from the summation formula for the spherical harmonics $Y_{lm}(\bf{\hat{r}})$ (see Appendix C for derivation)
\begin{multline}
\label{SumFormula}
\frac{4 \pi}{2l+1} \sum_{m=-l}^l Y_{lm}(\partial) Y_{lm}^\ast(\partial^\prime) = \\
= \sum_{k=0}^{\left[\frac{l}{2}\right]} c_{k} \left(\bm{\nabla} \cdot \bm{\nabla}^\prime \right)^{l-2k} \nabla^{2k} (\nabla^\prime)^{2k}.
\end{multline}
Here, $c_k$ are coefficients of the Legendre polynomial $P_l(x) = \sum_{k=0}^{\left[\frac{l}{2}\right]} c_{k} x^{l-2k}$, that are given by
\begin{equation}\
\label{ck}
c_k = \frac{(-1)^k (2l-2k)!}{2^l k! (l-k)!(l-2k)!},
\end{equation}
and $[x]$ denotes the largest integer, smaller or equal to $x$. In the following section we will apply representation \eqref{VpsStamp1} and summation formula \eqref{SumFormula} to derive relatively simple expression for the $p$- and $d$-wave zero-range potentials.

\subsection{Pseudopotentials for $p$ and $d$ wave in Cartesian coordinates}
\label{Sec:PseudoPD}

In the problems involving interacting atoms, very often an external potential does not posses a spherical symmetry, and the total angular momentum is not conserved. In such systems modeling of interactions in terms of pseudopotentials \eqref{Vps}, \eqref{VpsSt}, or \eqref{VpsStamp}, containing explicit integration with spherical harmonics, is not convenient. One of examples is the exact treatment of few-body systems confined in tight anisotropic harmonic trap,
that is relevant in context of experiments on ultracold atoms in optical lattices. In such cases, the pseudopotential \eqref{VpsStamp1} is far more easier to apply.

In this section we discuss pseudopotentials for $p$- and $d$-wave interactions, writing down explicitly the differential operators in the Cartesian coordinate system. Applying the summation formula \eqref{SumFormula} to the $p$-wave part of \eqref{VpsStamp1} we arrive at
\begin{equation}
\label{Vp}
V_{p}(\mbf{r}) = \frac{\pi \hbar^2 a_p^3}{\mu}
\stackrel{\leftarrow}{\nabla}
\delta(\mbf{r}) \frac{\partial^3}{\partial r^3} r^3 \stackrel{\rightarrow}{\nabla},
\end{equation}
where the symbol $\stackrel{\leftarrow}{\nabla}$
($\stackrel{\rightarrow}{\nabla}$) denotes the gradient operator
that acts to the left (right) of the pseudopotential, and $a_p$ is the $p$-wave energy-dependent scattering length: $a_p(k)^3 = - \tan \delta_1(k) /k^3$. This form of the pseudopotential with a scalar product of the two gradient operators have been first introduced by Omont~\cite{Omont}, and later by Kanjilal and Blume \cite{Kanjilal}, however, in the former the regularization operator was missing, whereas in the latter the regularization was different than ours.

Similar procedure applied to $d$-wave part leads to
\begin{align}
\label{Vd}
V_{d}(\mbf{r}) = \frac{\pi \hbar^2 a_d^5}{8\mu}
\sum_{i,j,k,l} & {D_{ijkl}}
 \!\!\!\!\! {\phantom{\bigg)}}^{\leftarrow} \!\!
%\bigg(\frac{\partial^2}{\partial x_i \partial x_j}\bigg) \delta(\mbf{r})
\Big(\partial^2_{x_i x_j}\Big) \delta(\mbf{r})
\frac{\partial^5}{\partial r^5} r^5
%\bigg(\frac{\partial^2}{\partial x_k \partial x_l}\bigg)^{\!\!\! \rightarrow}
\Big(\partial^2_{x_k x_l}\Big)^{\!\! \rightarrow}
\end{align}
where $D_{ijkl}= \delta_{ik} \delta_{jl} - \frac{1}{3} \delta_{ij} \delta_{kl}$ and
$a_d$ is the $d$-wave scattering length: $a_d(k)^5 = - \tan \delta_2(k)/k^5$. The zero-range potentials \eqref{Vp} and \eqref{Vd} presented here are slightly different from those introduced in Ref.~\cite{IdziaszekPRL}, differing in the form of the regularization potential and the position of the differential operator acting to the right. One can easily show that they are equivalent, whereas the present derivation is much easier to generalize to arbitrary partial wave, resulting in a formula \eqref{SumFormula} that can be used to generate pseudopotentials for $l > 2$.

Finally, we note that $s$-wave component of the pseudopotential in the representation \eqref{VpsStamp1} contains only the regularization operator $\partial_r r$, hence it does not require further transformations, and it can be directly applied in the problems with anisotropic external potentials \cite{IdziaszekPRA}. We stress that the use of $s$-wave zero-range potential without the projection operator is valid as long  as interactions takes place only in $s$ wave, and the radial wave function can exhibit only $1/r$ singular behavior at small $r$.

\section{Two atoms interacting in $p$-wave and confined in a harmonic trap potential}
\label{Sec:TwoAtoms}

\subsection{Analytical results for the zero-range potential}
\label{Sec:AnalRes}

In the following we consider two spin-polarized fer\-mions confined in an anisotropic harmonic trap. At low energies the interaction is dominated by the $p$-wave, and we model it using the zero-range potential (\ref{Vp}). We calculate the energy spectrum of the
system, that can be obtained solving the Lippmann-Schwinger equation for the wave function $\Psi(\mbf{r})$ of the relative motion
\begin{equation}
\label{GEq}
\Psi(\mbf{r}) = \int \! \! d^3r^\prime \, G(\mbf{r},\mbf{r}^\prime) V_p(\mbf{r}^\prime) \Psi(\mbf{r}^\prime),
\end{equation}
that for the discrete spectrum does not contain the source term. Here, $G_E(\mbf{r},\mbf{r}^\prime)= \langle \mbf{r}| (E-\hat{H})^{-1}| \mbf{r}^\prime \rangle$ is the single-particle
Green function, and $\hat{H}$ is the Hamiltonian including the external potential.
For anisotropic harmonic potential $G_E(\mbf{r},\mbf{r}^\prime)$ can be represented by the
following integral \footnote{Derivation is similar to the one presented in \cite{IdziaszekPRA}
for $G(0,\mbf{r})$}
\begin{align}
\label{Gho}
G_E(\mbf{r},\mbf{r}^\prime) = & \frac{(\eta_x \eta_y \eta_z)^{1/2}}{ (2 \pi)^{3/2} d^3 \hbar \omega}\
\int_{0}^{\infty}\!\!\! dt \, e^{E t/(\hbar \omega)} \nonumber \\
& \times \prod_{k} \frac{ \exp\left( \eta_k \frac{x_k x_k^\prime-\frac12(x_k^2+x_k^{\prime 2})
\cosh(t \eta_k)}{d^2 \sinh(t \eta_k)}
\right) }{\sinh(\eta_k t)^{1/2}},
\end{align}
which is convergent for energies smaller than the energy of zero-point oscillations
$E_0=\hbar (\omega_x + \omega_y + \omega_z)/2$. For $E>E_0$ the Green function can be determined by analytic continuation of (\ref{Gho}). Here, $d=\sqrt{\hbar/(\mu \omega)}$,
$\eta_k = \omega_k/\omega$, $\omega$ denotes some reference frequency, and product in (\ref{Gho}) runs over $k=x,y,z$. Inserting the pseudopotential (\ref{Vp}) into (\ref{GEq}), we obtain the following set of equations
\begin{equation}
\label{EqAC}
C_k = \frac{\pi \hbar^2 a_p^3}{\mu} \sum_{l} A_{kl}(E) C_l,
\end{equation}
where matrix elements $A_{kl}(E)$ are given by
\begin{equation}\label{Akl}
A_{kl}(E) = \left[\frac{\partial^{3}}{\partial r^{3}} r^{3} \frac{\partial^2}{\partial x_k \partial x_l^\prime} G_E(\mbf{r},\mbf{r}^\prime)\right]_{\mbf{r}=\mbf{r}^\prime=0},
\end{equation}
and coefficients $C_{l} \equiv \left[\partial_r^{3} r^{2} \partial_{x_l}  \Psi(\mbf{r})\right]_{\mbf{r}=0}$. The energy levels are determined from the secular equation derived from (\ref{EqAC}).

Let us analyze now the role of the regularization operator, to remove the singularities in derivatives of the Green function as $\mbf{r}, \mbf{r}^\prime \rightarrow 0$ in \eqref{Akl}. Performing the partial derivative with respect to $x_l^\prime$ and taking first the limit $\mbf{r}^\prime \rightarrow 0$ yields
\begin{multline}
\label{Gxl}
\left[\frac{\partial}{\partial x_l^\prime} G_E(\mbf{r},\mbf{r}^\prime)\right]_{\mbf{r}^\prime=0} \\
= - 2 \eta_l x_l \sqrt{\frac{\eta_x \eta_y \eta_z}{\pi^3}} \frac{e^{-\sum_k \eta_k x_k^2/2}}{d^5 \hbar \omega} H_l(\mbf{r}),
\end{multline}
where
\begin{align}\label{Hl}
H_l(E,\mbf{r}) = \int_{0}^{\infty}\!\!\! dt \, \frac{e^{t \left({\cal E}-\eta_l\right)}}{1-e^{-2t \eta_l}} \prod_{k} \frac{ \exp\left( - \eta_k x_k^2 \frac{e^{-2 t \eta_k}}{1-e^{-2 t \eta_k}}\right) }{(1-e^{-2 t \eta_k})^{1/2}},
\end{align}
and ${\cal E} = (E-E_0)/(\hbar \omega)$. The short-range singular behavior of $H_l(\mbf{r})$ is given by the small-$t$ asymptotics of the integrand in \eqref{Hl}. We calculate expansion of the integrand at $t=0$, and later subtract the diverging terms giving rise to the singularity of $H_l(\mbf{r})$ at $\mbf{r} \rightarrow 0$. This allows us to decompose $H_l(\mbf{r})$ into a regular part $F_l(\mbf{r})$ and the rest which is singular at $\mbf{r} \rightarrow 0$
\begin{equation}
H_l(\mbf{r}) = F_l(\mbf{r}) + \frac{\sqrt{\pi}}{4 \sqrt{\eta_x \eta_y \eta_z} \eta_l r^3} + \frac{(2 \varepsilon + \eta_x+\eta_y+\eta_z)\sqrt{\pi}}{8 \sqrt{\eta_x \eta_y \eta_z} \eta_l r},
\end{equation}
where
\begin{widetext}
\begin{align}\label{Fl}
F_l(E,\mbf{r}) = \int_{0}^{\infty}\!\!\! dt \, \left[\frac{e^{t \left({\cal E}-\eta_l\right)}}{1-e^{-2t \eta_l}} \prod_{k} \frac{ \exp\left( - \eta_k x_k^2 \frac{e^{-2 t \eta_k}}{1-e^{-2 t \eta_k}}\right) }{(1-e^{-2 t \eta_k})^{1/2}} - \frac{e^{-r^2/2t}}{\sqrt{32 \eta_x \eta_y \eta_z}\eta_l t^{5/2}} -
\frac{(2 \varepsilon + \eta_x+\eta_y+\eta_z)e^{-r^2/2t}}{\sqrt{128 \eta_x \eta_y \eta_z}\eta_l t^{3/2}}
\right],
\end{align}
\end{widetext}
With this result in mind one can now easily calculate $A_{kl}$. Calculating partial derivative of \eqref{Gxl} with respect to $x_k$, and later applying the regularization operator, we observe that the only nonvanishing terms comes from $F_l(r)$, whereas the singular terms in Eq.~\eqref{Hl} are removed by $\partial^{3}_r r^{3}$. This demonstrates that the regularization operator requires inclusion of the partial derivative of the form $\frac{\partial^{2l+1}}{\partial r^{2l+1}}$ in order to remove all the singular terms resulting from derivatives of the Green function. In this way we arrive at
\begin{equation}\label{Akl1}
A_{kl}(E) = - \frac{12 \eta_l}{d^5 \hbar \omega} \sqrt{\frac{\eta_x \eta_y \eta_z}{\pi^3}} F_l(E,0) \delta_{lk}
\end{equation}
where $F_l(0)$ is well defined, since the regularized integral \eqref{Fl} now converges in both integration limits when $\mbf{r}=0$ . Substituting \eqref{Akl1} into set of coupled equations \eqref{EqAC} determining the eigenenergies, we observe that they decouple, reducing to
\begin{equation}\label{Enk}
\frac{d^3}{\pi a_p^3} = - 12 \eta_k \sqrt{\frac{\eta_x \eta_y \eta_z}{\pi^3}} F_k(E,0).
\end{equation}
The energy spectrum generated by Eq.~\eqref{Enk} contains of three independent branches, that can labeled by $k = x,y,z$. This differs from the $s$-wave interacting atoms \cite{IdziaszekPRA}, where only one such a branch is found. We note that different branches do not describe to the excitations in the direction $k$, but rather results from the degeneracy of the $p$-wave states with respect to the projection of the angular momentum on the axis of quantization. We point out, that the total energy spectrum contains, apart from the eigenenergies determined from Eq.~\eqref{Enk}, also all the remaining eigenenergies of the noninteracting harmonic oscillator that have vanishing $p$-wave component at small $r$, and as a consequence are not affected by the $p$-wave pseudopotential.

In the following we focus on the case of an axially symmetric trap with $\omega_x = \omega_y = \omega_{\perp}=\eta \omega$ and $\omega_z = \omega$. In such a geometry the projection of the total angular momentum on the $z$-axis is conserved, and the different branches in \eqref{Enk} can be labeled by the projection quantum number $m$. First, we consider the case when $\eta$ ia an integer or an inverse of integer, when the integral defining $F_l(0)$ can be given in a closed form.

For pancake-shape traps, with $\eta=1/n$, the set of equations \eqref{Enk} reduces to
\begin{align}
\label{EnPan0}
\frac{d_z^3}{a_p^3} = & - \frac{8}{n} \sum_{k=0}^{n-1}
\frac{\Gamma(\frac{k+1/2}{n}-\frac{\cal E^\prime}{2}+ \frac 34)}
{\Gamma(\frac{k+1/2}{n}-\frac{\cal E^\prime}{2}- \frac 34)}, \quad m=0, \\
\label{EnPan1}
\frac{d_z^3}{a_p^3} = & - \frac{8}{n^2} \sum_{k,l=0}^{n-1}
\frac{\Gamma(\frac{k+l+1}{n}-\frac{\cal E^\prime}{2}+ \frac 14)}
{\Gamma(\frac{k+l+1}{n}-\frac{\cal E^\prime}{2}- \frac 54)}, \quad m=\pm 1,
\end{align}
where ${\cal E^\prime} = E/(\hbar \omega)$, $d_z=\sqrt{\hbar/(\mu \omega_z)}$,
$\Gamma(x)$ denotes the Euler Gamma function, and $m$ is the projection of the angular momentum on the $z$-axis. We note that in the presence of an axially symmetric trap the degeneracy between states with $m=0$ and $m=\pm1$ is lifted. In order to evaluate the integral \eqref{Fl} for $\eta=1/n$ we have applied the following identity $1/(1-e^{-x/n}) = \sum_{k=0}^{n-1} e^{-x k/n}/(1-e^{-x})$ (c.f. Ref.~\cite{IdziaszekPRA}).

The closed formulas can be also derived for cigar-shape traps with integer $\eta=n$:
\begin{widetext}
\begin{align}
\label{EnCig0}
\frac{d_z^3}{a_p^3} = & - 8 \frac{\Gamma(\frac 12-\frac{\cal E}{2})}{\Gamma(-1-\frac{\cal E}{2})}
- 6 \frac{\Gamma(\frac 12-\frac{\cal E}{2})}{\Gamma(1-\frac{\cal E}{2})} \left\{
-\frac{\cal E}{2} (n-1) +
\sum_{k=1}^{n-1} \frac{
F\left(1,\frac 12-\frac{\cal E^\prime}{2};1-\frac{\cal E^\prime}{2};e^{-i \frac{2\pi k}{n}}\right)}
{1 -e^{i \frac{2\pi k}{n}}} \right\} , \quad m=0, \\
\label{EnCig1}
\frac{d_z^3}{a_p^3} = & - 8 \frac{\Gamma(\frac 54-\frac{\cal E^\prime}{2})}{\Gamma(-\frac 14-\frac{\cal E^\prime}{2})}
- 6 \frac{\Gamma(\frac 54-\frac{\cal E^\prime}{2})}{\Gamma(\frac 74-\frac{\cal E^\prime}{2})}
\sum_{k=1}^{n-1} e^{-i \frac{2\pi k}{n}}
F\left(2,{\tst \frac 54-\frac{\cal E^\prime}{2}};{\tst \frac 74-\frac{\cal E^\prime}{2}};e^{-i \frac{2\pi k}{n}}\right),
\quad m=\pm 1,
\end{align}
\end{widetext}
where $F(a,b;c;x)$ denotes the hypergeometric function \cite{Abramowitz}. As can be easily verified, for particular case of spherically symmetric trap ($\eta=1$), Eqs. (\ref{EnPan0})-(\ref{EnCig1}) coincide with the result of Refs. \cite{Stock,Kanjilal}. To arrive at the Eqs.~\eqref{EnCig0}-\eqref{EnCig1} we have applied a similar technique as in \cite{IdziaszekPRA}, making use of the following identities: $1/(1-e^{-tn}) = \frac1n \sum_{k=0}^{n-1} (1-e^{-t-i 2 \pi k/n})^{-1}$, and $e^{-t(n-1)}/(1-e^{-tn})^2 = \frac{1}{n^2} \sum_{k=0}^{n-1} e^{-i 2 \pi k/n} /(1-e^{-t-i 2 \pi k/n})^2$.

In general, when $\eta$ is not an integer nor inverse of an integer, $F_l(E,0)$ can be calculated
numerically from the integral representation \eqref{Fl} for ${\cal E} <0$, where the integral converges. This yields the part of spectrum representing the bound states of the interaction potential. In analogy to $s$-wave interactions \cite{IdziaszekPRA}, for energies ${\cal E} >0$, above the zero point oscillation energy $E_0$ one can apply the following recurrence relations
that can be easily derived directly from the integral \eqref{Fl}
\begin{align}
F_z(E,0)-F_z(E-2\eta,0) & = 12 \eta \frac{\Gamma(\frac12-\frac{\cal E}{2})}{\Gamma(-\frac{\cal E}{2})} \\
F_x(E,0)-2 F_x(E-2\eta,0&) + F_x(E-4\eta,0)  \nonumber \\
 & = - 6 \eta^2 \frac{\Gamma(\frac{\eta}{2}-\frac{\cal E}{2})}{\Gamma(\frac{\eta}{2}+\frac12 -\frac{\cal E}{2})}.
\end{align}
The above equations together with \eqref{Fl} allows to evaluate $F_{x,y,z}(E,0)$ for arbitrary values of the energy and the trap anisotropy $\eta$.

\subsection{Energy spectrum for a model square-well interaction}
\label{Sec:SquareWell}

Results presented in the previous section have been derived using a zero-range potential, parameterized by the $p$-wave scattering volume $a_p^3$. For $s$-wave interactions, in the regime of ultracold collisions, the scattering cross-section acquires a constant value, and the pseudopotential can be described by a single, energy-independent parameter, e.g. $s$-wave scattering length. This however, does not hold when the interactions are dominated by the higher partial waves contributions, e.g. for polarized fermions, that can only interact via $p$-wave in the low-energy regime. In the case of $p$-wave interactions the range of applicability of the energy independent scattering length is very narrow \cite{IdziaszekPRL}, and typically one has to apply the energy-dependent pseudopotential with the scattering length determined by the phase shifts of the short range potential.

This behavior can be explained by analyzing the effective range expansion of the corresponding $p$-wave phase shift: $k^3 \cot \delta_1(k) = -1/a_p^3 - k^2/(2 R^{\ast}) + {\cal O}(k^4)$, where $R^{\ast}$ can be interpreted as an effective
range for $p$-wave interaction. The second term in the expansion can be neglected when $k |a_p| \ll (k R^{\ast})^{1/3}$, which combined with the condition for the
applicability of the pseudopotential: $k R^{\ast} \ll 1$, gives $k
|a_p| \ll 1$. Hence, the regime of applicability of the energy-independent $p$-wave pseudopotential is very limited, in contrast  to the $s$-wave interactions, where the
similar condition takes form $k |a_s| \ll (k R^{\ast})^{-1}$,
where $a_s$ is the $s$-wave scattering length and $(k R^{\ast})^{-1}$ is large.

In this section we consider a simple model of atoms interacting via square well potential. In the case of spherically symmetric trap the eigenenergies can be evaluated analytically, which allows to verify the accuracy of the pseudopotential treatment. For the energy-dependent pseudopotential, we evaluate the energy levels self-consistently from Eqs.~\eqref{EnPan0}-\eqref{EnCig1}, where $a_p^3$ is substituted with the energy scattering volume $a_p(k)^3 = - \tan \delta_1(k) /k^3$, where the $p$-wave phase shift $\delta_1(k)$ is evaluated for a square well potential. For negative energies this requires performing the analytical continuation of the phase shift, which can be obtained by substitution $k = i \kappa$ , with $E= -\hbar^2 \kappa^2/{2 \mu}$ \cite{StockTrapRes}.

\begin{figure}
   \includegraphics[width=8cm,clip]{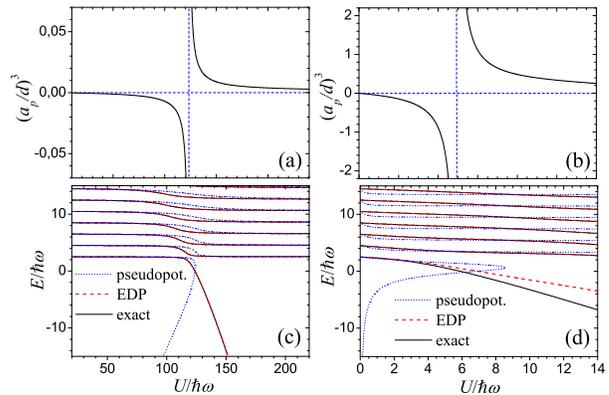}
	 \vspace{5mm}
	 \caption{
	 \label{fig:E100}
     (Color online) Upper panels: The scattering volume $a_p^3$ versus depth $U$ of the square well potential with diameter $R_0 = 0.2 d$ (left panel) and $R_0 = d$ (right panel). Lower panels: energy spectrum for two fermions interacting via the same square-well potentials, and confined in isotropic harmonic trap of frequency $\omega$, versus square-well depth $U$. The exact results (black solid) are compared with the zero-range potential model assuming energy-dependent (EDP, red dashed) and energy-independent (blue dotted) scattering length. Here, $d = \sqrt{\hbar / \mu \omega}$ denotes the harmonic oscillator length.
	 }
\end{figure}

Fig.~\ref{fig:E100} compares the exact energy levels for the square-well interaction, with the pseudopotential treatment assuming energy-dependent, and the standard energy-independent scattering volume. The comparison is performed for the two diameters of the square well: $R_0 = 0.2 d$, and
$R_0 = d$, where $d = \sqrt{\hbar / \mu \omega}$. The depth $U$ of the square well is varied in the vicinity of a resonance, that occurs when the bound state enters the potential well (c.f. the upper panels). We observe that EDP is very accurate, even when the size of the atom-atom interaction is comparable to the harmonic oscillator length determining the characteristic scale of the external potential, and some deviations can be observed only for deep bound states in the case of $R_0 = d$. On the other hand we note that the standard, energy-independent pseudopotential fails almost completely, and it works only very far from the resonance.

\begin{figure}
   \includegraphics[width=8.6cm,clip]{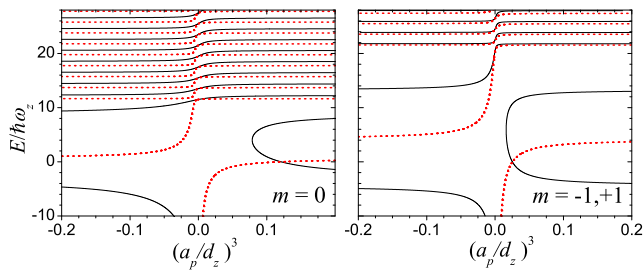}
	 \vspace{5mm}
	 \caption{
	 \label{fig:eta10}
     (Color online) Energy spectrum for two fermions interacting in $p$-wave and confined in axially symmetric harmonic trap with $\eta = \omega_{\perp}/\omega_z = 10$, versus the scattering volume $a_p^3$ scaled by the harmonic oscillator length $d_z = \sqrt{\hbar / \mu \omega_z}$.
     The left and right panels show the eigenenergies for different values of the $z$-component of the angular momentum. The black-solid lines presents the spectrum calculated assuming the energy independent scattering volume, whereas the red-dotted lines are obtained for energy-dependent scattering volume $a_p^3(k)$ of the square-well potential with $R_0 = 0.2 d_z$ (see text for details).
	 }
\end{figure}

\begin{figure}
   \includegraphics[width=8.6cm,clip]{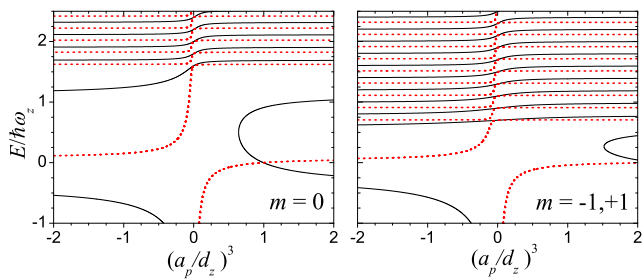}
	 \vspace{5mm}
	 \caption{
     \label{fig:eta01}
	 (Color online) Same as Fig.~\ref{fig:eta10} but for $\eta = 0.1$.
	 }
\end{figure}

Having established the regime of applicability of the pseudopotential method, we turn now to the case of anisotropic traps, where the exact levels are more difficult to calculate, and we perform the analysis solely within the pseudopotential treatment. Figs.~\ref{fig:eta10} and \ref{fig:eta01} shows the energy levels calculated for $\eta=10$ and
$\eta=1/10$. In contrast to the previous analysis, here we plot the results as a function of the $p$-wave scattering volume, determining it for each depth $U$ of the square-well at $E = 0$. In the case of energy-independent pseudopotentials this corresponds exactly to the energies given by Eqs~\eqref{EnPan0}-\eqref{EnCig1}, respectively. This way of presenting the results let us to analyze the origin of the loop-like structures that can be observed in the part of the spectrum describing the bound states for calculations with the energy-independent pseudopotential. We observe that such structures are artificial, and disappear completely when one resorts to the correct, energy-dependent pseudopotential. Similarly as in the case of isotropic trap, we observe that predictions of energy-dependent and energy-independent pseudopotentials differ almost completely, except the vicinity of points when $a_p^3 = 0$.

%%%%%%%%%%%%%%%%%% Figure 1 %%%%%%%%%%%%%%%%%%%%%%
%%%%%%%%%%%%%%%%%%%%%%%%%%%%%%%%%%%%%%%%%%%%%%%%%%%

%%%%%%%%%%%%%%%%%% Figure 2 %%%%%%%%%%%%%%%%%%%%%%
%%%%%%%%%%%%%%%%%%%%%%%%%%%%%%%%%%%%%%%%%%%%%%%%%%%

%%%%%%%%%%%%%%%%%% Figure 3 %%%%%%%%%%%%%%%%%%%%%%
%%%%%%%%%%%%%%%%%%%%%%%%%%%%%%%%%%%%%%%%%%%%%%%%%%%

\subsection{Energy spectrum in the vicinity of a $p$-wave Feshbach resonance}
\label{Sec:Feshbach}

In this section we investigate the energy spectrum of two trapped spin-polarized fermions, with interactions modified by means of a $p$-wave magnetic Feshbach resonance. We start with deriving a formula describing the energy-dependent $p$-wave scattering volume $V(k) = - \tan \delta_1(k) /k^3$ for a Feshbach resonance. To this end we utilize the multichannel quantum-defect theory (MQDT) \cite{Seaton,Greene,Mies}, which has proven very effective as
few-parameter approach for describing scattering and bound states in neutral-neutral \cite{Gao,Burke}, charged-neutral \cite{Watanabe,IdziaszekIon}, and charged-charged \cite{Seaton} particle collisions.

In brief, MQDT is based on the separation of the
short-range and the long-range behavior of the scattering wave function. The short-range wave function is insensitive to the total energy $E$, and can be described with a few quantum-defect parameters, that are insensitive both to the energy and the relative angular-momentum of the colliding particles \cite{GaoMQDT,IdziaszekIon}. On the contrary,
at large distances the wave function vary rapidly with $E$, together with other scattering quantities defined at $r \rightarrow \infty$ (e.g. phase shifts).

In our treatment of a $p$-wave resonance, we consider for simplicity only one open and one closed channel. This is similar to CI (configuration interaction) model of a Feshbach resonance \cite{MiesFeshbachCI}, where a single closed channel represents the effects of all the closed channels contributing to a resonance. In the two channel case, the application of MQDT is
straightforward, and after some simple algebra (see for instance \cite{MiesJulienne}), one obtain the following phase shift $\delta(E,l)$ in the open channel \cite{JulienneFeshbachMQDT}
\begin{equation}
\label{ResPhase}
\delta(E,l) = \delta_\mrm{bg}(E,l) - \arctan \left( \frac{\frac\Gamma 2 C^{-2}(E,l)}{E-E_{n}+\frac \Gamma 2 \tan \lambda(E,l)}\right).
\end{equation}
The first term $\delta_\mrm{bg}(E,l)$ is the phase shift resulting from the scattering in the open channel only, called the background phase shift. The second term describes the resonant contribution originating from a bound state in the closed channel, with energy $E_{n}$ located close to the threshold of the open channel. The width of the resonance $\Gamma$ is multiplied by the MQDT function $C^{-2}(E)$ that accounts for a proper threshold behavior as $k\rightarrow0$. In MQDT $C^{-2}(E)$ relates the amplitudes of the solutions with short-range and long-range WKB-like normalization \cite{Mies,Julienne:89}. The second MQDT function, $\tan \lambda(E)$, describes a mixing between the two linearly independent solutions that occur in the quantum regime when we transform from the short-range to the long-range normalized wave functions \cite{Mies,Julienne:89}. When $k\rightarrow0$ and the semiclassical approximation become valid at all distances, we have $\tan \lambda(E) \rightarrow 0$ and $C^{-2}(E) \rightarrow 1$, and the result \eqref{ResPhase} reduces to the Fano-Beutler formula for the phase shift near a resonance \cite{FanoRau}.

We can assume that the energy of a bound state, varies approximately linearly with the magnetic field $B$:
\begin{equation}
\label{En}
E_n(B) = \delta\mu (B -B_n),
\end{equation}
where $B_n$ is the magnetic field when the bound state crosses the threshold of the open channel, and $\delta\mu$ is the difference of magnetic moments of the open and closed  channels. In the case of a power law interaction potential $r^{-s}$, the phase shift and MQDT functions exhibit the following Wigner threshold behavior \cite{MiesRaoult}
\begin{align}
\label{Thr1}
\delta_\mrm{bg}(E) &\stackrel{E \rightarrow 0}{\longrightarrow} A_l k^{2l+1} &\quad 2l+1 \leq s-2 \\
C^{-2}(E) &\stackrel{E \rightarrow 0}{\longrightarrow} B_l k^{2l+1} &\quad \textrm{all l} \\
\label{Thr3}
\tan \lambda(E) &\stackrel{E \rightarrow 0}{\longrightarrow} \tan \lambda(0) &\quad \textrm{all l}
\end{align}
The threshold behavior given by \eqref{Thr1}-\eqref{Thr3} is determined by the long-range part of the interaction potential, and for van der Waals forces the higher order terms in $k$ can be neglected when energy is smaller than some characteristic energy $E_\mrm{vdW}$ associated with $C_{6}/r^6$ potential. For alkali  $E_\mrm{vdW}$ ranges from $0.1$mK to $30$mK \cite{JulienneFeshbachMQDT}, hence in the ultracold regime ($E \lesssim 1\mu$K), one can safely use the approximations \eqref{Thr1}-\eqref{Thr3}. Making use of Eqs.~\eqref{Thr1}-\eqref{Thr3} we can introduce the width of magnetic Feshbach resonance $\Delta B$:
\begin{equation}
\label{DB}
\lim_{E \rightarrow 0} \frac{\Gamma}{2} \frac{C^{-2}(E)}{\tan \eta_{bg}(E)} = - \delta \mu \, \Delta B,
\end{equation}
and the resonance position $B_0$, that is shifted from $B_n$ due to the coupling between the open and closed channel
\begin{equation}
\label{B0}
B_0 = B_n + \frac{\Gamma}{2 \delta \mu} \lim_{E \rightarrow 0} \tan \lambda(E).
\end{equation}
Now, substituting Eqs.~\eqref{En}, \eqref{DB}, \eqref{B0} into \eqref{ResPhase}, and performing some simple algebra we arrive at
\begin{equation}
\label{TanD}
\begin{split}
\tan \delta(E) & = \tan \delta_\mrm{bg}(E) \\
& \times \left(1 - \frac{\Delta B \left[1+ \tan^{2} \delta_\mrm{bg}(E)\right]}{B-B_0 - \frac{E}{\delta \mu} + \Delta B \tan^{2} \delta_\mrm{bg}(E)} \right)
\end{split}
\end{equation}
For $s$-wave scattering, formula \eqref{TanD} yields the expression for the energy-dependent scattering length (Eq.(75) in Ref.~\cite{IdziaszekPRA}), that apart from the scattering in very tight trapping potentials, reduces to the standard expression $a(B) = a_\mrm{bg}\left[1-\Delta B/(B-B_0)\right]$. In contrast, for $p$-wave resonances it is important to
take into account the energy-dependence of the resonance position and width. In this case the energy-dependent scattering volume $V(E)$ reads
\begin{equation}
\label{VE}
V(E) = V_\mrm{bg}(E) \left[1 - \frac{\Delta B \left(1+  \frac{E^3}{E_\mrm{bg}^3}\right)}{B-B_0 - \frac{E}{\delta \mu} + \Delta B \frac{E^3}{E_\mrm{bg}^3}} \right]
\end{equation}
Here, $E_\mrm{bg}= \hbar^2/(2 \mu a_\mrm{bg})$ is the energy associated with background scattering length $a_{bg} = \lim_{E \rightarrow 0} V_\mrm{bg}(E)^{1/3}$. Typically $E_\mrm{bg} \sim E_\mrm{vdW}$, except the case when the background scattering volume is large: $V_\mrm{bg} \gg R_\mrm{vdW}$, compared with the van der Waals length
$R_\mrm{vdW} = \frac{1}{2}(2 \mu C_6/\hbar^2)^{1/4}$. When $E_\mrm{bg} \sim E_\mrm{vdW}$, and for $E \ll E_{vdW}$, when one does not need to take into account the effective-range expansion of $V_\mrm{bg}(E)$, Eq.~\eqref{VE} can be simplified to
\begin{equation}
\label{VEappr}
V(E) \cong V_\mrm{bg} \left[1 - \frac{\Delta B}{B-B_0 - E/\delta \mu} \right]
\end{equation}
We observe that in comparison to $s$-wave resonances, the position of the resonance is shifted for finite energies of colliding particles, and as we show later, this shift is significant for atoms confined in the tight traps ($\omega \gtrsim 1 $kHz).

Application of the result \eqref{VEappr} in the Eqs.~\eqref{EqAC} and \eqref{Akl} yields the energy spectrum of the two spin-polarized fermions versus magnetic field. We have calculated the energy levels for a specific example of $^{40}$K atoms, where $p$-wave Feshbach resonances has been observed experimentally and studied theoretically. In Ref.~\cite{Ticknor} Feshbach resonances have been parameterized in the sprit of an effective-range expansion
\begin{equation}
\label{CotD}
k^3 \cot \delta_1(k,B) = - \frac{1}{v(B)} + \frac{k^2}{R(B)}
\end{equation}
The zero-energy scattering volume $v$, and the effective range $R$ were approximated by the second-order polynomials in $B$
\begin{align}
\label{vB}
\frac{1}{v(B)} & = c^{(0)} + c^{(1)} B + c^{(2)} B^2 \\
\label{RB}
\frac{1}{R(B)} & = d^{(0)} + d^{(1)} B + d^{(2)} B^2
\end{align}
where the expansion coefficient have been determined by fitting to the multichannel numerical calculations, remaining with well agreement with the experimental measurements~\cite{Ticknor}. The expansion coefficients, listed in Table \ref{Tab1}, depend on the projection of the relative angular momentum, $m_l$, on the axis of the magnetic field. This multiplet structure of a $p$-wave Feshbach resonance results from the small anisotropic spin dipole-dipole interaction, that couples different partial waves, and breaks the rotational invariance.
%%%%%%%%%%%%%%%%%% Table 1 %%%%%%%%%%%%%%%%%%%%%%
\begin{table}
\begin{ruledtabular}
\begin{tabular}{cccc}
$|m_l|$ &
$\begin{array}{cc} c^{(0)} \textrm{(a.u.)} \\ d^{(0)} \textrm{(a.u.)} \end{array}$  &
$\begin{array}{cc} c^{(1)} \textrm{(a.u.)} \\ d^{(1)} \textrm{(a.u.)} \end{array}$  &
$\begin{array}{cc} c^{(2)} \textrm{(a.u.)} \\ d^{(2)} \textrm{(a.u.)} \end{array}$  \\
\hline \hline \\
0 &
$\begin{array}{ll} \phantom{-} 8.68155\times10^{-5} \\ -1.64805 \end{array}$  &
$\begin{array}{ll} -8.29778\times10^{-7} \\ \phantom{-} 0.01523 \end{array}$  &
$\begin{array}{rr} 1.97732\times10^{-9} \\ -3.54471\times10^{-5} \end{array}$ \\ \hline \\
1 &
$\begin{array}{ll} \phantom{-} 7.83424\times10^{-5} \\ -2.36792 \end{array}$  &
$\begin{array}{ll} -7.45662\times10^{-7} \\ \phantom{-} 0.02264 \end{array}$  &
$\begin{array}{rr} 1.76807\times10^{-9} \\ -5.45051\times10^{-5} \end{array}$ \\
\end{tabular}
\end{ruledtabular}
\caption{\label{Tab1} Expansion coefficients of Eqs.~\eqref{vB} and \eqref{RB} for a $p$-wave Feshbach resonance between two $^{40}$K atoms in $\left|f=\frac{9}{2},m_f=-\frac{7}{2}\right\rangle$ state at $B_0 \approx 198\mrm{G}$ determined in Ref.~\cite{Ticknor}. The coefficients depend on the projection quantum number $m_l$ of the relative angular momentum along the axis of magnetic field.
}
\end{table}
%%%%%%%%%%%%%%%%%%%%%%%%%%%%%%%%%%%%%%%%%%%%%%%%%%%

We have fitted expansion \eqref{CotD} to our formula \label{VE1}, in the energy range $0 \leq E \leq 500$kHz and magnetic field range $195 \mrm{G} < B < 205 \mrm{G}$, obtaining the resonance parameters listed in Table~\ref{Tab2}. The agreement between \eqref{CotD} and \eqref{VEappr} remained on the level of few percent, which was similar to the accuracy of the effective range formula
\eqref{CotD}, when fitting to the numerical calculations in Ref.~\cite{Ticknor}. We have also tried the fit to the more accurate expression \eqref{VE}, including the effective range expansion for $V_\mrm{bg}(E)$, but this has not further improved the quality of the fit.
%%%%%%%%%%%%%%%%%% Table 3 %%%%%%%%%%%%%%%%%%%%%%
\begin{table}
\begin{ruledtabular}
\begin{tabular}{ccccc}
$|m_l|$ & $B_0$[G] & $\Delta B$[G] & $\delta\mu/h$ [kHz/G] & $a_{bg}$[$a_0$] \\
\hline
$0$ & $198.85$ & $-20.342$ & $93.093$ & $-104.26$ \\
$1$ & $198.37$ & $-22.956$ & $92.667$ & $-99.539$
\end{tabular}
\end{ruledtabular}
\caption{\label{Tab2} Resonance position $B_0$, resonance width $\Delta B$, the difference of magnetic moments $\delta\mu$ and the background scattering length $a_\mrm{bg} = (V_\mrm{bg})^{1/3}$, for a $p$-wave Feshbach resonance between two $^{40}$K atoms in $\left|f=\frac{9}{2},m_f=-\frac{7}{2}\right\rangle$ state, depending on their relative-angular-momentum projection quantum number $m_l$ along the axis of magnetic field. The numbers have been extracted from the data in \cite{Ticknor} by fitting to the resonance formula \eqref{VEappr}.
}
\end{table}
%%%%%%%%%%%%%%%%%%%%%%%%%%%%%%%%%%%%%%%%%%%%%%%%%%

\begin{figure}
   \includegraphics[width=8cm,clip]{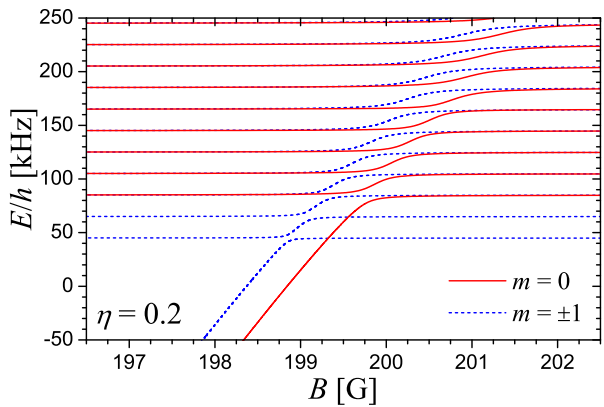}
	 \caption{
	 \label{fig:FPan}
     (Color online) Energy spectrum of two interacting fermions in the vicinity of a $p$-wave Feshbach resonance, versus magnetic field $B$ aligned along $z$ direction. The atoms are confined in an axially-symmetric trap with $\eta = \omega_\perp/\omega_z = 1/5$ and $\omega_z = 50$~kHz. The blue-solid and red-dashed lines correspond to different projections of the angular momentum along $z$ axis.
	 }
\end{figure}

\begin{figure}
   \includegraphics[width=8cm,clip]{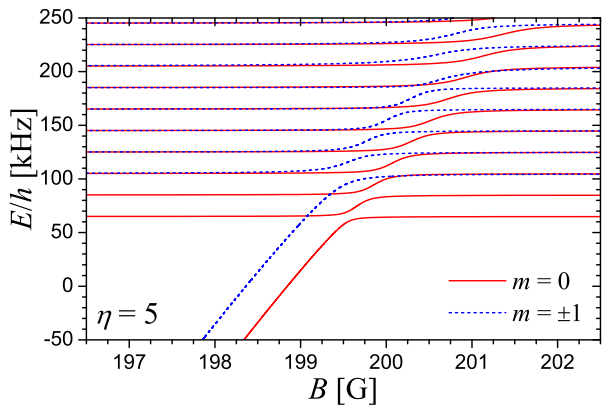}
	 \caption{
	 \label{fig:FCig}
	 Same as Fig.~\eqref{fig:FPan} but for a cigar-shape trap with $\eta = \omega_\perp/\omega_z = 5$.
	 }
\end{figure}

Figures \ref{fig:FPan} and \ref{fig:FCig} presents the energy spectrum for two $^{40}$K atoms, in the vicinity of a $p$-wave resonance at $B_0 \approx 198 G$. For simplicity we have chosen direction of applied magnetic field along the $z$-axis, in order to preserve the axial symmetry of the system. In such geometry, the eigenstates with different $m_l$ are not coupled, and the resonance occurs at different value of magnetic field for $m_l =0$ and  $|m_l| = 1$, respectively. The resonance can be observed as avoided crossing between the trap states represented by the horizontal lines, and the bound state corresponding to the skew line in the spectrum. The effects of the trap anisotropy can be observed, as the different width of avoided
crossing, depending on the excitation quantum number of the trap state (c.f. the avoided crossings with the 5th and 6th trap excited state for $|m_l|=1$ in Figure \ref{fig:FCig}). We note that similarly to the square-well model, the inclusion of the energy-dependent scattering volume, eliminates the presence of the loop-like structures in the energy spectrum below the zero-point-oscillations energy.
%%%%%%%%%%%%%%%%%% Figure 4 %%%%%%%%%%%%%%%%%%%%%%
%%%%%%%%%%%%%%%%%%%%%%%%%%%%%%%%%%%%%%%%%%%%%%%%%%%
%%%%%%%%%%%%%%%%%% Figure 5 %%%%%%%%%%%%%%%%%%%%%%
%%%%%%%%%%%%%%%%%%%%%%%%%%%%%%%%%%%%%%%%%%%%%%%%%%%

%\begin{equation}
%\begin{split}
%\tan \, & \delta(E,L)  = \tan \delta_\mrm{bg}(E,l) \\
%& \times \left(1 - \frac{\Gamma}{2} \frac{C^{-2}(E)(\tan \delta_\mrm{bg} + \tan^{-1} \delta_\mrm{bg})}{E-E_n + \frac{\Gamma}{2} \tan \lambda + \frac{\Gamma}{2} \tan \delta_\mrm{bg} C^{-2}(E)} \right)
%\end{split}
%\end{equation}

\section{Summary and conclusions}
\label{Sec:Summary}

Summarizing, we have presented analytical solutions for two spin-polarized fermions interacting in $p$-wave and confined in a harmonic potential of arbitrary symmetry. For certain class of axially symmetric traps, with the aspect ratio $\eta$ integer or inverse of an integer, we have derived closed formulas for the energy spectrum. In general the eigenenergies can be easily evaluated numerically from some simple integral for energies smaller than the zero-point-oscillations energy: $E<E_0$, or with the help of recurrence relation for $E>E_0$.

As an intermediate step we have discussed in details derivation of a generalized pseudopotential for all partial waves, introduced in Ref.~\cite{IdziaszekPRL}. In the present paper we use slightly different form of the $p$-wave zero-range potential in Cartesian coordinates, that turned out to be more convenient for our analytical calculations. In addition, we have presented the way to generate similar pseudopotentials for higher partial waves, showing the result also for the $d$-wave pseudopotential. Because of several different formulations of the zero-range potentials for $\ell > 0$ existing in the literature, we have discussed how our pseudopotential connects to other approaches, and we have specified the conditions when different pseudopotentials are equivalent.

The higher partial wave interactions are vastly different from the $s$-wave one, and in the ultracold regime they typically require application of an energy-dependent pseudopotential (EDP). We illustrated this first for a model square-well potential, comparing the pseudopotential method with exact energy levels. We observed that EDP works well even when the range of atom-atom interaction is comparable to the characteristic size of an external potentials, which is well beyond the applicability of an ordinary pseudopotential. Then, we considered interactions of a pair of atoms close to a $p$-wave Feshbach resonance. In the framework of the multichannel quantum-defect theory, we have derived relatively simple formula for the scattering volume in the vicinity of a resonance, that is quite similar to its $s$-wave analogue, which is well known in the literature. In this  way we obtained the energy spectrum of two trapped fermions near a $p$-wave Feshbach resonance, that contrary to predictions based on the energy-independent pseudopotential, does not contain characteristic loop structures in the spectrum. In fact, the magnetic field dependence of the spectrum is quite similar to the one for $s$-wave interactions, in the presence of tight confining potentials \cite{Bolda,IdziaszekPRA}.

We stress that our approach allows in principle for determination of eigenstates, however the resulting wave functions would exhibit the short-range $1/r^3$ divergent behavior, and thus they could not be normalized in the standard way. The remedy is the application a delta-shell potential \cite{Stock}, that introduces a short-range cut-off that can be associated with the range of the atom-atom interaction. Another solution is based on the regularization of the scalar product, according to the theory of generalized functions, or equivalently introducing a modified scalar product to define the normalized wave-functions \cite{Pricoupenko2}. Moreover, for narrow $p$-wave Feshbach resonances the contribution of the closed channel to the total wave function is significant, and cannot be neglected in the normalization. We stress that MQDT allows for rigorous determination of closed-channel component, which will not affect the results presented here for the spectrum, nevertheless should be taken into account while determining eigenstates. These extensions, however, are beyond the scope of the present paper focusing solely on the properties of the energy spectrum, and will be investigated elsewhere.

\begin{acknowledgments}
The author acknowledge helpful discussion with Paul S. Julienne. This work was supported by the Polish Government Research Grant for years 2007-2009.
\end{acknowledgments}

\appendix

\section{Hadamard finite part regularization}
\label{Sec:App1}

In this section we present a reasoning that leads to a finite part regularization of Eq.~\eqref{Pf}. We basically follow the approach presented in the work of Estrada and Kanwal \cite{Estrada}. Let's consider the following integral that diverges at the lower limit of integration
\begin{equation}
F(\varepsilon) = \int_{\varepsilon}^{\infty} \ud x \, \frac{\phi(x)}{x^k}.
\end{equation}
At small $\varepsilon$, the integral exhibit the following behavior
\begin{equation}
\label{Fexp}
F(\varepsilon) = F_0(\varepsilon) + c_0 \ln \varepsilon + \sum_{j=1}^{k-1}, \frac{c_j}{\varepsilon^j}
\end{equation}
with some constants $c_j$, and $F_0(\varepsilon)$ that is regular in the limit $\varepsilon \rightarrow 0$. The finite part in the sense of Hadamard is defined as (see e.g. \cite{Kanwal})
\begin{equation}
\mrm{Pf} \left(\int_{\varepsilon}^{\infty} \ud x \, \frac{\phi(x)}{x^k} \right)= \lim_{\varepsilon \rightarrow 0^+} F_0(\varepsilon).
\end{equation}
Splitting the integration into two intervals: $(\varepsilon,a)$ and $(a,\infty)$ with some arbitrary $a>\varepsilon$, we can extract terms that diverge in the way described in Eq.~\eqref{Fexp}
\begin{align}
\int_{\varepsilon}^{\infty} \!\!\ud x \, & \frac{\phi(x)}{x^k}  = \nonumber \\
& = \int_{a}^{\infty}  \!\!\ud x \, \frac{\phi(x)}{x^k} + \int_{\varepsilon}^{a}  \frac{\ud x}{x^k} \left[ \phi(x) - \sum_{j=0}^{k-1} \frac{\phi^{(j)}(0)}{j!} x^j \right] \nonumber \\
& \quad + \int_{\varepsilon}^{a} \!\!\ud x \, \sum_{j=0}^{k-1} \frac{\phi^{(j)}(0)}{j!} x^{j-k} \\
& =  \int_{a}^{\infty}  \!\!\ud x \, \frac{\phi(x)}{x^k} + \int_{\varepsilon}^{a}  \frac{\ud x}{x^k} \left[ \phi(x) - \sum_{j=0}^{k-1} \frac{\phi^{(j)}(0)}{j!} x^j \right] \nonumber \\
& \quad + \sum_{j=0}^{k-2} \frac{\phi^{(j)}(0)}{j!(j-k+1)}\left( a^{j-k+1} - \varepsilon^{j-k+1} \right) \nonumber \\
& \quad + \frac{\phi^{(k-1)}(0)}{(k-1)!}\left( \ln a - \ln \varepsilon \right).
\end{align}
By comparison with \eqref{Fexp}, the finite part in the sense of Hadamard reads
\begin{align}
\mrm{Pf} \Bigg(&\int_{0}^{\infty} \!\!\ud x \, \frac{\phi(x)}{x^k} \Bigg)  =  \nonumber \\
& = \int_{a}^{\infty}  \!\!\ud x \, \frac{\phi(x)}{x^k} + \int_{0}^{a}  \frac{\ud x}{x^k} \left[ \phi(x) - \sum_{j=0}^{k-1} \frac{\phi^{(j)}(0)}{j!} x^j \right] \nonumber \\
& \quad + \sum_{j=0}^{k-2} \frac{\phi^{(j)}(0)}{j!(j-k+1)} a^{j-k+1} + \frac{\phi^{(k-1)}(0)}{(k-1)!} \ln a . \nonumber
\end{align}
Substituting $a = \varepsilon$ and taking limit $\varepsilon \rightarrow 0^+$ we obtain
\begin{multline}
\label{Pfa}
\textrm{Pf} \left(\frac{1}{x^k}\right) = \lim_{\varepsilon \rightarrow 0^+} \Bigg[ \frac{H(r-\varepsilon)}{x^k}
+ \sum_{j=0}^{k-2} \frac{(-1)^j \delta^{(j)}(x)}{j! (j-k+1) \varepsilon^{k-j-1}} \\
+ \frac{(-1)^{k-1} \ln \varepsilon \delta^{(k-1)}(x)}{(k-1)!} \Bigg]
\end{multline}
Here, $H(x)$ denotes the Heaviside function
\begin{equation}
H(x) = \begin{cases}
0 & x<0 \\
1 & x>0
\end{cases}
\end{equation}

\section{Expressing the pseudopotential in terms of $Y_{lm}(\partial)$ operator}

In order to prove identity \eqref{Ident1} we calculate the action of the distribution $T = \delta^{(l)}(r) Y_{lm}(\mbf{\hat{r}})/r^2$ on a test function $\phi$ that is regular and differentiable to arbitrary order
\begin{equation}
\label{B1}
\left\langle \phi, T\right\rangle = \int_0^\infty \!\! \ud r \, \delta^{(l)}(r) f_{lm}(r),
\end{equation}
where
\begin{equation}
f_{lm}(r) \equiv  \int \!\! d\Omega \, Y_{lm}(\hat{\mbf{r}}) \phi(\mbf{r}).
\end{equation}
Performing Taylor expansion of $\phi(\mbf{r})$ around $\mbf{r} = 0$, one can easily obtain the small-$r$ behavior of $f_{lm}(r)$
\begin{equation}
\label{TaylorExp}
f_{lm}(r) = c_{lm} r^l + {\cal O} (r^{l+1}), \quad r \rightarrow 0,
\end{equation}
with
\begin{align}
c_{lm} & = \frac{1}{l!} \int \!\! \ud\Omega \, Y_{lm}(\hat{\mbf{r}})
\left((\hat{\mbf{r}} \nabla)^l \phi \right)(0)
\end{align}
In the next step we substitute \eqref{TaylorExp} into Eq.~\eqref{B1} and make use of the property \eqref{OrthoProperty}
\begin{align}
\left\langle \phi, T \right\rangle & =
(-1)^l  l! c_{lm} \nonumber \\
& = \frac{4 \pi (-1)^l l!}{(2l+1)!!} \left[ Y_{lm}(\partial)\sum_{l^\prime m^\prime} c_{l^\prime m^\prime} r^{l^\prime} Y_{l^\prime m} (\mbf{r}) \right]_{\mbf{r} = 0} \nonumber\\
& = \frac{4 \pi (-1)^l l!}{(2l+1)!!} \left[ Y_{lm}(\partial)\phi(\mbf{r}) \right]_{\mbf{r} = 0}
\end{align}
This finally yields
\begin{align}
\left\langle \phi, T \right\rangle  = \frac{4 \pi l!}{(2l+1)!!}
\left\langle \phi, Y_{lm}(\partial) \delta(\mbf{r})\right\rangle,
\end{align}
which is equivalent to \eqref{Ident1}.

\section{Summation formula for $Y_{lm}(\partial)$ operators}
\label{Sec:App3}

The proof of Eq.~\eqref{SumFormula} can be conveniently conducted using the Fourier transformation. We start with
\begin{multline}
\label{SumFormula1}
\sum_{m=-l}^l Y_{lm}(\partial) Y_{lm}^\ast(\partial^\prime)  f(\mbf{r}) g(\mbf{r}^\prime)= \\
= \sum_{m=-l}^l \! Y_{lm}(\partial) Y_{lm}^\ast(\partial^\prime) \! \iint\!\frac{d^3k \,d^3q}{(2 \pi)^6} \tilde{f}(\mbf{k}) \tilde{g}(\mbf{q}) e^{-i (\mbf{k} \mbf{r} + \mbf{q} \mbf{r}^\prime)}
\end{multline}
where $f(\mbf{r})$ and $g(\mbf{r}^\prime)$ are some test functions with Fourier transforms
\begin{align}
\tilde{f}(\mbf{k}) & = \int \!\! \frac{d^3r}{(2 \pi)^3} \, e^{i \mbf{k} \mbf{r}} f(\mbf{r}), \\
\tilde{g}(\mbf{k}) & = \int \!\! \frac{d^3r}{(2 \pi)^3} \, e^{i \mbf{k} \mbf{r}} g(\mbf{r}).
\end{align}
The differential operator acting on the exponential function produces: $Y_{lm}(\partial) e^{-i \mbf{k} \mbf{r}} = (-i)^l Y_{lm}(\hat{\mbf{k}}) k^l$. Substituting this result into \eqref{SumFormula1} and making use of the standard summation formula for the spherical harmonics
(see for instance \cite{Abramowitz})
\begin{equation}
\sum_{m=-l}^l Y_{lm}(\hat{\mbf{k}}) Y_{lm}^\ast(\hat{\mbf{q}}) =
P_l(\hat{\mbf{k}}\hat{\mbf{q}}),
\end{equation}
we arrive at
\begin{multline}
\label{SumFormula2}
\sum_{m=-l}^l Y_{lm}(\partial) Y_{lm}^\ast(\partial^\prime)  f(\mbf{r}) g(\mbf{r}^\prime)= \\
= \frac{2l+1}{4 \pi (-1)^l} \! \iint\!\frac{d^3k \,d^3q}{(2 \pi)^6} P_l(\hat{\mbf{k}}\hat{\mbf{q}}) k^l q^l \tilde{f}(\mbf{k}) \tilde{g}(\mbf{q}) e^{-i (\mbf{k} \mbf{r} + \mbf{q} \mbf{r}^\prime)}.
\end{multline}
Substituting the explicit form of the Legendre polynomial $P_l(x) = \sum_{k=0}^{\left[l/2\right]} c_{k} x^{l-2k}$, with $c_k$ defined in \eqref{ck}, we obtain
\begin{multline}
\label{SumFormula3}
\sum_{m=-l}^l Y_{lm}(\partial) Y_{lm}^\ast(\partial^\prime)  f(\mbf{r}) g(\mbf{r}^\prime)= \\
= \sum_{k=0}^{\left[\frac{l}{2}\right]} c_{k} \left(\bm{\nabla} \cdot \bm{\nabla}^\prime \right)^{l-2k} \nabla^{2k} (\nabla^\prime)^{2k} f(\mbf{r}) g(\mbf{r}^\prime).
\end{multline}

\bibliography{pwave}

\end{document}